\documentclass[twocolumn]{revtex4}
\usepackage{graphicx}
\newcommand{\e}{{\rm e}}
\newcommand{\ep}{\varepsilon}

\newcommand{\nn}{\nonumber}
\newcommand{\la}{\label} 

\newcommand{\be}{\begin{equation}}
\newcommand{\ee}{\end{equation}}
\newcommand{\ba}{\begin{eqnarray}}
\newcommand{\ea}{\end{eqnarray}}

\begin{document}
\title{A Fundamental Theorem on the Structure of Symplectic Integrators}

\author{Siu A. Chin}

\affiliation{Department of Physics, Texas A\&M University,
College Station, TX 77843, USA}

\begin{abstract}
I show that the basic structure of symplectic integrators is governed
by a theorem which states {\it precisely}, how symplectic integrators 
with positive coefficients cannot be corrected beyond second order. 
All previous known results can now be derived quantitatively from this theorem. 
The theorem provided sharp bounds on second-order 
error coefficients explicitly in terms of factorization coefficients. By 
saturating these bounds, one can derive fourth-order 
algorithms analytically with arbitrary numbers of operators. 

\end{abstract}
\maketitle

\section {Introduction}

Algorithms for solving diverse physical problems ranging
from celestial mechanics\cite{yoshi,mcl02,chinchen03,chinsante}, 
quantum statistical mechanics\cite{ti,li,jang,chincor,crb} to quantum 
dynamics\cite{shen,serna,chinchen01,chinchen02} can all be derived
from approximating the evolution operator ${\rm e}^{\ep(T+V)}$ 
in the product form
\ba
&&\prod_{i=1}^N
\e^{t_i\ep T}\e^{v_i\ep V}
=\exp\biggl[\ep\bigl( e_T T+e_V V+\ep e_{TV}[T,V]\nonumber\\
&&+\,\ep^2 e_{TTV}[T,[T,V]]+\ep^2 e_{VTV}[V,[T,V]]+..\bigr)\biggr]
\label{prodform} 
\ea
with factorization coefficients $\{t_1\}$
and $\{v_i\}$. Classically, every product of the form (\ref{prodform})
produces a symplectic integrator for integrating classical equations
of motion. For solving {\it time-irreversible} problems involving the 
diffusion operator, such as the quantum statistical 
trace\cite{ti,li,jang,chincor,crb} or the imaginary time Schr\"odinger 
equation\cite{fchinm,auer,ochin,chinkro}, 
one must also insists that these coefficients be positive.
Since $T$ and $V$ are non-commuting operators, the general
problem of deriving approximations of the form
(\ref{prodform}) beyond second order, regardless of the sign of 
the coefficients, is extremely difficult. For $N>3$, most higher 
order algorithms can only be found by using symbolic algebra and numerical 
methods\cite{mcl02,chinchen02,kos96,ome02,ome03}. In this work, 
I prove a fundamental theorem relating the error coefficients
$e_{TV}$, $e_{TTV}$ and $e_{VTV}$ from which
one can deduce previous known results quantitatively  
and construct fourth-order algorithms analytically 
with arbitrary $N$.  
 
The error coefficients
$e_T$, $e_V$, $e_{TV}$, $e_{TTV}$, and $e_{VTV}$ in (\ref{prodform})
are related to the
factorization coefficient $\{t_i\}$ and $\{v_i\}$ via\cite{chincor}
\ba
e_T=\sum_{i=1}^N t_i,&&\quad e_V=\sum_{i=1}^N v_i,
\la{prim}\\
\frac12+e_{TV}&=&\sum_{i=1}^N t_i u_i\,,
\label{etv} \\
{1\over 6}+{1\over2}e_{TV}+e_{TTV}
&=&{1\over2}\sum_{i=1}^N t_i(s_i+s_{i-1}) u_i\,,
\label{ettv}\\
{1\over 6}+{1\over 2}e_{TV}-e_{VTV}
&=&{1\over 2}\sum_{i=1}^N t_i u_i^2\,,
\label{evtv}
\ea
in terms of useful variables
\be s_i=\sum_{j=1}^i t_j\,,\quad u_i=\sum_{j=i}^N v_j\,. 
 \label{si} 
\ee 
Satisfying the primary constraints
$e_T=1$ and $e_V=1$ implies that $s_N=1$ and 
\be
u_1=1.
\la{uone}
\ee
For $\{t_i\}>0$, (\ref{evtv}) is a quadratic form in $u_i$
whose minimum can be determined 
subject to linear constraints (\ref{etv}) and (\ref{ettv}). 
This then leads to an inequality relating $e_{TV}$, $e_{TTV}$, 
and $e_{VTV}$ sufficient to prove that the general product (\ref{prodform})
cannot be corrected (to be explained below) beyond second order with positive 
coefficients\cite{chincor}. However, the exact minimum
was not determined because the required sum, first appeared in Suzuki's 
work\cite{suzukinogo}, 
\be
g=\sum_{i=1}^N s_is_{i-1}(s_i-s_{i-1})={1\over 3}(1-\delta g),
\label{gsec}
\ee
was not known in closed form. The inequality was therefore weak,
excluding the possibility of being equal. 
Surprisingly, a closed form exists
and was found in Ref.\cite{nosix},
\be
\delta g=\sum_{i=1}^N t_i^3.
\label{delg}
\ee
The minimum was then determined, but without explicitly incorporating 
the constraint (\ref{uone}). Recently, it was realized\cite{chinfour} that
constraint (\ref{uone}) can be enforced without affecting any
of the equations (\ref{etv}), (\ref{ettv}) and (\ref{evtv}) 
by simply setting $t_1=0$. The resulting
minimum is then only true for algorithms whose first operator is $\e^{v_1\ep V}$.
Since classically this operator updates the velocity (momentum) variable,
the constraint (\ref{uone}) dictates that the minimum derived in Ref.\cite{nosix}
only holds for {\it velocity}-type algorithms. By interchanging
$T\leftrightarrow V$ and $\{t_i\}\leftrightarrow \{v_i\}$ in
all of the above, the constraint $e_T=1$ now dictates that $v_1=0$ and 
another minimum holds for {\it position} type 
algorithm whose first operator is $\e^{t_1\ep T}$. One is finally able 
to state the exact relationship between $e_{TV}$, $e_{TTV}$, and $e_{VTV}$
directly in terms of {\it either} $\{t_i\}$ {\it or} $\{v_i\}$
corresponding to either velocity or position-type algorithms.
By constructing integrator whose error coefficients are precisely
at the quadratic minimum, the condition for being fourth-order 
can be directly stated, and easily solved for, in terms of $\{t_i\}$ 
{\it or} $\{v_i\}$. One is then able to construction fourth-order
integrators analytically for arbitrary $N$ as it was done in Ref.\cite{chinfour}.
The current theorem provided sound theoretical support and unified
derivation of results obtained in Ref.\cite{chinfour}.
    
\section {The Theorem}

The constrained minimum of the quadratic 
form in (\ref{evtv}) can be obtained by the method of Lagrange multiplier.
Since this has been worked out in details in Ref.\cite{chincor}
(but for a much weaker goal), we will just summarize the results. 
For $t_1=0$ and $\{t_{i>1}\}>0$, minimize
\be F = {1\over 2}\sum_{i=1}^N t_i u_i^2
-\lambda_1 \left( \sum_{i=1}^N t_i u_i\right) -\lambda_2 \left(
\sum_{i=1}^N t_i(s_i+s_{i-1}) u_i\right ) \ee 
with respect to $u_i$ gives, 
\be
u_i=\lambda_1+\lambda_2(s_i+s_{i-1}). \label{usol} 
\ee
Substituting this back to satisfy constraints (\ref{etv}) and
(\ref{ettv}) determines $\lambda_1$ and $\lambda_2$: 
\be
\lambda_1+\lambda_2=\frac12+e_{TV},
\la{lsum} 
\ee 
\be \lambda_1+\lambda_2+g\lambda_2=
{1\over 3}+e_{TV}+2 e_{TTV},
\la{lsum2}
\ee 
where $\sum_{i=1}^N
t_i(s_i+s_{i-1})^2=1+g $.
The minimum of the quadratic form
is then \ba
F_{min}&=&{1\over 2}\sum_{i=1}^N t_i [\lambda_1+\lambda_2(s_i+s_{i-1})]^2,\nonumber\\
 &=&{1\over 2}[ (\lambda_1+\lambda_2)^2+g\lambda_2^2],\nonumber\\
 &=&{1\over 2}({1\over 2}+e_{TV})^2+{1\over{2g}}(2\, e_{TTV}-{1\over 6})^2.
\label{min}
\ea
Setting the LHS of (\ref{evtv}) greater or {\it equal} (this is the most important point,
the main contribution of this work) to $F_{min}$ gives,
\smallskip

\noindent
{\bf Theorem, Part A:}
For $t_1=0$ and $\{t_{i>1}\}>0$, the error coefficients
for the product of operators in (\ref{prodform}) obey the inequality,  
\be
e_{VTV}\le\frac1{24}-\frac12 e_{TV}^2-\frac6{1-\delta g}
\left (e_{TTV}-\frac1{12}\right)^2,
\la{tha}
\ee
or, after slight arrangement,
\ba
\e_{VTV}&+&\frac12 e_{TV}^2-e_{TTV}\nn\\
&\le&-\frac1{24}\delta g
-\frac6{1-\delta g}(e_{TTV}-\frac{1}{12}\delta g)^2,
\la{notcora}
\ea
where $\delta g$ is given by (\ref{delg}). Since
$0<\delta g<1$ for $\{t_{i>1}\}>0$ and $e_T=1$, the second form
shows that the LHS of (\ref{notcora}) is strictly negative.
Note that $t_1=0$ does not prevent the algorithm from
being completely general. Nothing stops us from
considerating algorithms with $v_1=0$, in which case,
the result will be a position-type algorithm. This part of the theorem
simply regard $\{t_i\}$ as independent variables. 

By interchanging
$T\leftrightarrow V$ and $\{t_i\}\leftrightarrow \{v_i\}$ in (\ref{prodform}),
the error coefficients changes respectively,
$\e_{TV}\rightarrow -\e_{TV}$, $\e_{TTV}\rightarrow -\e_{VTV}$ and 
$\e_{VTV}\rightarrow -\e_{TTV}$. Making the substitution in (\ref{tha})
yields,
\smallskip

\noindent
{\bf Theorem, Part B:}
For $v_1=0$, and $\{v_{i>1}\}>0$, the error coefficients
for the product of operators
\ba
&&\prod_{i=1}^N
\e^{v_i\ep V}\e^{t_i\ep T}
=\exp\biggl[\ep\bigl( e_T T+e_V V+\ep e_{TV}[T,V]\nonumber\\
&&+\,\ep^2 e_{TTV}[T,[T,V]]+\ep^2 e_{VTV}[V,[T,V]]+..\bigr)\biggl].
\label{prodform2} 
\ea
obey the inequality,  
\be
e_{TTV}\ge-\frac1{24}+\frac12 e_{TV}^2+\frac6{1-\delta g^\prime}
\left (e_{VTV}+\frac1{12}\right)^2,
\la{thb}
\ee
or in the form
\ba
\e_{TTV}&-&\frac12 e_{TV}^2-e_{VTV}\nn\\
&\ge&\frac1{24}\delta g^\prime
+\frac6{1-\delta g^\prime}(e_{VTV}+\frac{1}{12}\delta g^\prime)^2,
\la{notcorb}
\ea
where the corresponding $\delta g^\prime$ is given by
\be
\delta g^\prime=\sum_{i=1}^N v_i^3.
\label{delgp}
\ee
Again (\ref{notcorb}) shows that the LHS is strictly positive.
We will regard (\ref{notcora}) and (\ref{notcorb}) as fundamental
statements of our theorem. To explain this, we need to mention 
symplectic corrector (or process) algorithms\cite{wis96,mar97,blan99}.

If $\rho$ denotes an approximation to $\e^{\ep(T+V)}$ of the 
product form (\ref{prodform}),
then  $\rho$ is ``correctable" if 
\be
\rho^\prime=S\rho S^{-1}
\ee
is correct to higher-order in $\ep$ for some operator $S$ also of the
general form (\ref{prodform}) but with no sign restriction on 
its factorization coefficients\cite{wis96,mar97,blan99}.  
If $\rho$ is correctable, then its 
trace, equal to the trace of $\rho^\prime$, will be correct to higher
order in $\ep$. This is important for calculating the quantum statistical 
trace of an approximate density matrix, as in path integral
Monte Carlo calculations\cite{ti,li,jang,chincor,crb}. The criterion for $\rho$ 
to be correctable to at least third-order in $\ep$ is\cite{chincor}
\be
\e_{VTV}+\frac12 e_{TV}^2-e_{TTV}=0
\la{corcond}
\ee
However, if $\{t_i\}\ge 0$, then (\ref{notcora}) shows that this is not possible.
And if $\{v_i\}\ge 0$, then (\ref{notcorb}) also shows that this is not
possible. Our theorem states {\it precisely}, how forward symplectic
integrator of the product form, consisting of only operators $T$ and $V$, 
{\it cannot} be corrected beyond second order. 

A much weaker form of this theorem, that the LHS of (\ref{corcond}) 
cannot be zero, has been proved previously by Chin\cite{chincor}, and 
by Blanes and Casas\cite{blanes05} using a very different method. The current 
theorem is much sharper, stating the precise amount by which the correctability 
condition (\ref{corcond}) is being 
missed, when $\{t_i\}\ge 0$, and when $\{v_i\}\ge 0$.   

Two main corollaries: {\bf 1)} It is easy to force $e_{TV}=0$; all
odd-order error terms will vanish if we simply choose factorization 
coefficients that are left-right symmetric in (\ref{prodform}) or 
(\ref{prodform2}). If $e_{TV}=0$, then the correctability criterion is just 
$e_{TTV}=e_{VTV}$. However, (\ref{notcora}) and (\ref{notcorb})
both show that there is an unbridgeable gap between the two coefficients;
they can never be equal. In particular, they can never {\it both} equal to zero.
This corollary is the Sheng-Suzuki Theorem\cite{sheng,suzukinogo}: 
there cannot be factorization algorithms of the form (\ref{prodform}) 
with positive coefficients beyond second order. Again our current theorem is 
more quantitative in showing that if $\{t_i\}>0$, then the gap is given 
by (\ref{notcora}) and if if $\{v_i\}>0$, then the gap is given by (\ref{notcorb}).
{\bf 2)} If both $e_{TV}$ and
$e_{TTV}$ are zero, then (\ref{notcora}) implies that 
\be
e_{VTV}\le-\frac1{24}\frac{\delta g}{1-\delta g},
\la{evtvb}
\ee
and can only vanish if $\delta g=0$, requiring at least
one $t_i$ to be negative. If both $e_{TV}$ and
$e_{VTV}$ are zero, then (\ref{notcorb}) implies that 
\be
e_{TTV}\ge\frac1{24}\frac{\delta g^\prime}{1-\delta g^\prime},
\la{ettvb}
\ee
and can only vanish if $\delta g^\prime=0$, requiring at least
one $v_i$ to be negative. This corollary is the Goodman-Kaper
theorem\cite{goldman}: beyond second order, factorization algorithms 
of with only operators $T$ and $V$  must have at 
least a pair of negative coefficients ($t_k$, $v_k$). Our current theorem is 
again much more quantitative with symmetric forms (\ref{evtvb}) 
and (\ref{ettvb}).

\section {Constructing Fourth-Order Algorithms }

Since all odd-order error terms vanish with left-right
symmetric coefficients, fourth-order algorithms can be
obtained by forcing both $e_{TTV}$ and $e_{VTV}$ to zero.
Let's consider first velocity-type algorithms described by
Part A of the theorem. When $e_{TV}$ and $e_{TTV}$ are both zero, 
the bound for $e_{VTV}$ (\ref{evtvb}) is the actual error 
coefficient for algorithms with $u_i$ given by
(\ref{usol}), corresponding to
\be
v_i=-\lambda_2(t_i+t_{i+1}),
\la{vi}
\ee
\be
v_1=\frac12+\lambda_2(1-t_2)\,\,\,{\rm and}\,\,\,
v_N=\frac12+\lambda_2(1-t_N),
\la{vn}
\ee
with $\lambda_2$ given by (\ref{lsum}) and (\ref{lsum2}),
\be
\lambda_2=-\frac12\frac1{1-\delta g}.
\ee
Eq.(\ref{vi}) is true for all algorithms whose quadratic form
is {\it stationary} with respect to $u_i$. Thus the equal sign
in (\ref{evtvb}) holds even for negative $t_i$.
A fourth-order algorithm results if we choose a left-right
symmetric set of $\{t_i\}$ with $t_1=0$ such that $e_T=1$
and $\delta g=0$. For example, for $N=6$, we can choose
$t_6=t_2$, $t_5=t_3$. The constraints
\ba
2t_2+2t_3+t_4&=&1\nn\\ 
2t_2^3+2t_3^3+t_4^3&=&0
\la{supfor}
\ea
can be solved by setting $t_2=\alpha t_3$, giving
\be
t_4=-2^{1/3}\left(1+ \alpha^3 \right)^{1/3} t_3,
\ee
\be
t_3=\frac1{2\left(1+\alpha\right)
-2^{1/3}\left( 1+ \alpha^3 \right)^{1/3}}.
\ee
The case of $\alpha=0$ reduces back to the well known
Forest-Ruth integrator\cite{for90}. For $\delta g=0$, coefficients 
$v_i$ given by (\ref{vi})-(\ref{vn}) are {\it linearly} 
related to $\{t_i\}$. For position-type algorithms, one
can simply exchange operators $T\leftrightarrow V$
and their coefficients $\{t_i\}\leftrightarrow \{v_i\}$.
For further examples of constructing this type of algorithms, 
see Ref.\cite{chinfour}.

Instead of forcing $e_{VTV}$ to vanish on the RHS of
(\ref{prodform}), one can simply move the entire
operator $$\exp(\ep^3e_{VTV}[V,[T,V]])$$
back to the LHS, and combine the commutator  
$\ep^3e_{VTV}[V,[T,V]]$ symmetrically with
one or more operator $\ep v_i V$. For $T=p^2/2$, 
both classically\cite{chin} and quantum mechanically\cite{suzuvtv}, 
$[V,[T,V]]$ corresponds to an additional {\it gradient} force or 
potential similar to $V$. By doing this, one gets around the Sheng-Suzuki 
theorem in producing fourth-order {\it forward} ($t_{i>1}>0)$ symplectic 
integrator by factorizing $\e^{\ep(T+V)}$ through an additional 
operator $[V,[T,V]]$. This results in a far richer family of algorithms 
since any set of symmetric coefficient $\{t_{i>1}\}$ (regardless of sign) 
satisfying $e_T=1$ will now yield a fourth-order algorithm.
For example, for $N=4$, taking $t_2=t_3=t_4=1/3$ produces
\be
v_1=v_4=\frac18,\quad v_2=v_3=\frac38,\quad e_{VTV}=-\frac1{192}, 
\la{alg4d}
\ee
which is forward algorithm 4D\cite{chinchen01}. But one can also
take $t_2=t_4=2/3$, $t_3=-1/3$, giving 
\be
v_1=v_4=\frac18,\quad v_2=v_3=\frac38,\quad e_{VTV}=-\frac5{96}. 
\la{alg4dp}
\ee
This also illustrates that the Goodman-Kaper theorem no longer
holds if one includes $[V,[T,V]]$ in the factorization process. 
More examples of deriving velocity-type gradient algorithms 
are given in Ref.\cite{chinfour}.  

To construct position-type algorithms, we invoke Part B of the
theorem. Since we have a preference for keeping the commutator $[V,[T,V]]$,
we must set $e_{TV}=e_{TTV}=0$ in (\ref{thb}) and solve for
$e_{VTV}$,
\be
\e_{VTV}=-\frac1{12}(1-\sqrt{1-\delta g^\prime}),
\la{epos}
\ee
where we have picked the solution which vanishes with $\delta g^\prime$. 
Under the interchange
$T\leftrightarrow V$ and $\{t_i\}\leftrightarrow \{v_i\}$, 
the corresponding equation (\ref{lsum2}) for $\lambda_2$ reads
\be
\frac12 +g^\prime \lambda_2=\frac13-2 e_{VTV},
\ee
from which one deduces
\be
\lambda_2=-\frac12\frac1{\sqrt{1-\delta g^\prime}}.
\la{lmpos}
\ee   
Thus any set of symmetric $\{v_i\}$ with $v_1=0$ and $\e_V=1$ will produce
a fourth-order algorithm via
\be
t_i=-\lambda_2(v_i+v_{i+1}),
\la{ti}
\ee
\be
t_1=\frac12+\lambda_2(1-v_2)\,\,\,{\rm and}\,\,\,
t_N=\frac12+\lambda_2(1-v_N),
\la{tn}
\ee
with $\lambda_2$ and $e_{VTV}$ given by (\ref{lmpos}) and (\ref{epos}).
For $N=4$, $v_2=v_3=v_4=1/3$, this gives
\be
t_1=t_4=\frac12 \Bigl(1-\frac1{\sqrt{2}}\Bigr),\quad
t_2=t_3=\frac1{2\sqrt{2}},
\ee
and
\be
e_{VTV}=-\frac1{12}\Bigl(1-\frac23 \sqrt{2}\Bigr).
\ee
More examples of deriving position-type forward algorithms
can be found in Ref.\cite{chinfour}.

In conclusion, I have presented a fundamental theorem on symplectic 
integrators from which fourth-order algorithms of arbitrary length can 
be constructed for solving diverse physical problems.
Other important properties of symplectic integrators can also
be deduced from this theorem.

I thank Drs. Blanes and Casas for a comment on Ref.\cite{nosix} which
triggered my understanding on the subject. 
This work is supported, in part, by a National Science Foundation 
grant, No. DMS-0310580.

\newpage
\centerline{REFERENCES}

\end{document}